\documentclass[fleqn,preprint,12pt]{elsarticle_arXiv08110943v4}

\renewcommand{\theequation}{\arabic{section}.\arabic{equation}}

\journal{Nucl. Phys. B 815 (2009) 198}
\renewcommand{\today}{arXiv:0811.0943v4 [hep-th]}

\usepackage{graphicx}
\usepackage{bm}
\usepackage{mathrsfs}
\usepackage[latin1]{inputenc}
\usepackage{stmaryrd}
\usepackage{array}
\usepackage{psfrag}
\usepackage{dsfont}
\usepackage{epsfig}
\usepackage{titletoc}
\usepackage{float}   
\usepackage{wrapfig} 
\usepackage{amsmath}

\def\be {\begin{equation}}
\def\ee {\end{equation}}
\def\ba {\begin{eqnarray}}
\def\ea {\end{eqnarray}}

\newcommand{\equivdef}{\equiv}
\newcommand{\bra}[1]{\left(#1\right)}

\newcommand{\fr}[2]{{\frac{#1}{#2}}}

\newcommand{\vierbein}[2]{e_{#1}^{\;\;#2}}
\newcommand{\inversevb}[2]{e^{#1}_{\;\;#2}}

\begin{document}

\begin{frontmatter}

\title{Lorentz violation and black-hole thermodynamics}

\date{\versiondate}

\author{G. Betschart} \ead{betschart@particle.uni-karlsruhe.de}
\author{E. Kant} \ead{kant@particle.uni-karlsruhe.de}
\author{F.R. Klinkhamer\corref{cor1}}
\cortext[cor1]{Corresponding Author}
\ead{frans.klinkhamer@physik.uni-karlsruhe.de}
\address{Institute for Theoretical Physics, University of Karlsruhe (TH),
         76128 Karlsruhe, Germany}

\begin{abstract}
We consider nonstandard photons from nonbirefringent modified Maxwell theory
and discuss
their propagation in a fixed Schwarzschild spacetime background. This
particular modification of Maxwell theory is Lorentz-violating and allows for
maximal photon velocities differing from the causal speed $c$
of the asymptotic background spacetime.
In the limit of geometrical optics, light rays from modified Maxwell theory
are found to propagate along null geodesics in an effective metric. We observe that
not every Lorentz-violating theory with multiple maximal velocities different
from the causal speed $c$  modifies the notion of the event horizon, contrary
to naive expectations. This result implies that not every
Lorentz-violating theory with multiple maximal velocities necessarily leads
to a contradiction with the generalized second law of thermodynamics.
\end{abstract}
\begin{keyword}
      \hspace*{-3mm}Lorentz violation,\! geometrical optics,\! black-hole thermodynamics
\PACS 11.30.Cp          \sep  42.15.-i            \sep 04.70.Dy
\end{keyword}
\end{frontmatter}

\newpage
\setcounter{equation}{0}
\section{Introduction}

Recently, Dubovsky and Sibiryakov~\cite{DubovskySibiryakov2006}
investigated a specific Lorentz-violating theory, based on the
ghost-condensate  model of Ref.~\cite{ArkaniHamed-etal2004}, which allows
for different maximal velocities of different species of particles in a
preferred reference frame.  The authors of
Ref.~\cite{DubovskySibiryakov2006} demonstrated that this, in turn, implies
the possibility of constructing a perpetuum mobile of the second kind
(which relies on heat transfer from a cold body to a hot body, without other changes).
The basic idea is that different
particles in the theory considered have a different notion of the event
horizon of a Schwarz\-schild black hole.\footnote{Recall that, in a
Lorentz-invariant theory, the standard Schwarzschild horizon occurs at
$r=2\, G_\text{N}M/c^2$, where $c$ is the \emph{unique} maximal
attainable velocity of all particles.  Naively, Lorentz-noninvariant
particles with $v_\text{max}< c$ would then have an effective horizon
\emph{outside} the one of the Lorentz-invariant
particles.\label{ftn:naive-horizon}} Making use of the
quantum-mechanical Hawking effect,
the conclusion is that particles with different maximal velocities
measure different effective black-hole masses and temperatures, which,
in principle, allows for the construction of a perpetuum
mobile~\cite{DubovskySibiryakov2006}.

Subsequently, Eling \emph{et al.}~\cite{Eling-etal2007} proposed a
classical mechanism for Lorentz-violating theories with multiple propagation
speeds, which also indicates a conflict between the generalized second
law of thermodynamics and such Lorentz-violating theories.
(The suggested mechanism~\cite{Eling-etal2007} is similar to Penrose's
energy-extraction mechanism for a rotating black hole, which relies
on the existence of the so-called ergosphere region. For the nonrotating
black hole in the Lorentz-violating theory considered there is also
an accessible ergosphere region.)
Eling \emph{et al.} \cite{Eling-etal2007} and Jacobson and Wall
\cite{JacobsonWall2008} further suggest that the violation of the
generalized second law of
thermodynamics~\cite{Bekenstein1974,Hawking1975,UnruhWald1983,FrolovPage1993}
may be a \emph{general} feature
of Lorentz-symmetry-breaking theories with different maximal velocities
for different species of particles.

However, the argument of Eling \emph{et al.} \cite{Eling-etal2007} is limited
to Lorentz-violating theories, for which test particles
propagating in the background Schwarz\-schild spacetime
(with mass parameter $M$) perceive effective metrics which are
again simply  Schwarz\-schild but with different masses and horizons. In this
article, we will consider a particular Lorentz-violating modification of Maxwell
theory, which allows for a broader class of effective metrics.
We introduce an additional free
parameter ``four-vector'' $\xi^\mu$ which determines the nature of the
resulting effective metric for the photon field.

It turns out that the effective metric
from parameters $\xi^\mu$ is not necessarily a Schwarz\-schild
metric with a different mass.
Other possibilities are the original metric itself (i.e., unchanged $M$)
or even a non-Ricci-flat metric with an event horizon
at the original Schwarz\-schild radius $r=2\, G_\text{N}M/c^2$.
Since the latter possibilities  exclude the
construction of the particular type of perpetuum mobile
considered in Refs.~\cite{DubovskySibiryakov2006,Eling-etal2007},
we conclude that not every Lorentz-violating theory
with different maximal velocities of different particle species
necessarily leads to a manifest contradiction with black-hole thermodynamics.

Our main result agrees with that of Sagi and
Bekenstein~\cite{SagiBekenstein2007} obtained in a different theory.
These authors considered black-hole solutions
in a Lorentz-violating tensor-vector-scalar theory of gravity, which
implies different propagation velocities of light and gravitational waves.
In this particular tensor-vector-scalar theory of gravity,
the construction of a classical perpetuum mobile along the lines of
Ref.~\cite{Eling-etal2007} can also be excluded and the violation of the
second law as suggested in Ref.~\cite{DubovskySibiryakov2006} can be
avoided if the conjectured equality of effective graviton
radiation temperature and photon Hawking temperature holds.
The Lorentz violation studied in the present article is of a different
type and allows for different maximal velocities between photons and
matter, in contrast to the tensor-vector-scalar theory of gravity,
where all matter propagation is described by the same physical metric.

The possibility of constructing Lorentz-violating theories which do not
contradict the generalized second law of
thermodynamics~\cite{Bekenstein1974,Hawking1975,UnruhWald1983,FrolovPage1993}
may suggest a super selection rule for the Lorentz-violating parameters
of an explicitly Lorentz-violating theory,
as it seems reasonable to assume that a theory must not allow
for the construction of a perpetuum mobile. (For us,  an ``explicitly
Lorentz-violating theory''  is a theory where the Lorentz-violating
parameters already occur at the level of the action, such as
in the Standard Model Extension action of Ref.~\cite{ColladayKostelecky1998}.)
However, the coupling of explicitly
Lorentz-violating theories to gravity remains an open problem
\cite{Kostelecky2004}, which must be solved before we can fully
understand the interplay of Lorentz violation and curved spacetime
and are able to draw definitive conclusions on the subject.

\setcounter{equation}{0}
\section{Theory: General aspects}\label{sec:Theory-General}

\subsection{Units and conventions}

Electromagnetism (standard and nonstandard)
is, first, described with rationalized MKSA units
and, then, with natural units in order to get $c=G_\text{N}=k_\text{B}=\hbar =1$.
Spacetime indices are denoted  by Greek letters and correspond to
the standard spherical coordinates $t,r,\theta, \phi$,
while local Lorentz indices are denoted by Latin letters and run from $0$ to
$3$. The symbol $\eta_{\mu\nu}$ stands for the flat-spacetime Minkowski metric
and $g_{\mu\nu}$ for the
curved-spacetime metric, both with signature $\left(+,-,-,-\right)$.
The absolute value of the determinant of the metric is abbreviated
as $g=|\,\text{det}\,g_{\mu\nu}|\,$. The vierbein is denoted
$e^\mu_{\;\;a} = g^{\mu\nu}\eta_{ab}\,\vierbein{\nu}{b}$ and obeys the
relations $\inversevb{\mu}{a}\vierbein{\mu}{b} = \delta_a^{~b}$,
$\inversevb{\mu}{a}\vierbein{\nu}{a} = \delta_{~\nu}^{\mu}$,
 and $g_{\mu\nu}=\vierbein{\mu}{a}\vierbein{\nu}{b}\,\eta_{ab}$.
Finally, covariant derivatives are written as $D_\mu$.

\subsection{Nonbirefringent modified Maxwell theory in flat spacetime}

Modified Maxwell theory in flat spacetime is a generalized $U(1)$ gauge theory
with a Lagrange density which consists of the standard Maxwell term and an
additional Lorentz-violating bilinear term.
Specifically, the flat-spacetime photonic Lagrange density reads:
\begin{equation}\label{eq:LagMMflat}
\mathcal{L}_\text{modMax, flat}=
-\frac{1}{4} F_{ab}F_{cd}\,\eta^{ac}\eta^{bd}
-\frac{1}{4}\kappa^{abcd}F_{ab}F_{cd}\,,
\end{equation}
in terms of the standard Maxwell field strength
$F_{ab}\equiv \partial_a A_b -\partial_b A_a$.
The Lorentz-violating ``tensor'' $\kappa^{abcd}$
has the same symmetries as the Riemann curvature tensor,
as well as a double-trace condition:
\begin{equation}\label{eq:kappa-abcd}
\kappa^{abcd} = \kappa^{[ab]\,[cd]}, \quad \kappa^{abcd} =
\kappa^{cdab},\quad \kappa^{ab}_{~~ab} = 0\ .
\end{equation}
Under the simplest assumption usually discussed in the literature,
$\kappa^{abcd}$ is constant and has 19 independent parameters.
More generally, $\kappa^{abcd}(x)$
could be an arbitrary, but fixed, ``tensor field,''
effectively corresponding to more than 19 parameters,
possibly an infinite number.

The following \emph{Ansatz} \cite{BaileyKostelecky2004} reduces
modified Maxwell theory to the nonbirefringent sector:
\begin{equation}\label{eq:ansatznonbire}
\kappa^{abcd}=\frac{1}{2}
\left(\eta^{ac}\widetilde{\kappa}^{bd}-\eta^{ad}\widetilde{\kappa}^{bc}
     +\eta^{bd}\widetilde{\kappa}^{ac}-\eta^{bc}\widetilde{\kappa}^{ad}\right),
\end{equation}
in terms of a symmetric and traceless matrix $\widetilde{\kappa}^{ab}$.
Later on, it will be convenient to employ the following
decomposition of $\widetilde\kappa^{ab}$:
\begin{subequations}\label{eq:ansatzxi}
\begin{eqnarray}
\widetilde{\kappa}^{ab}
&=&
\kappa\left(\xi^a\xi^b-\eta^{ab}\, \xi^c\xi_c/4\right),
\label{eq:Ansatzxi-kappa-tensor}
\\[1mm]
\kappa
&\equivdef&
\frac{4}{3}\,\widetilde\kappa_{ab}\,\xi^a\xi^b,
\label{eq:Ansatzxi-kappa-scalar}
\end{eqnarray}
\end{subequations}
relative to a normalized parameter four-vector $\xi^a$ with
$\xi_a \xi^a=1$ or $-1$, corresponding to the timelike or spacelike case,
respectively. The discussion of
geometrical optics for modified Maxwell theory to be given in
Sec.~\ref{sec:Effective-metric} is valid for general parameter
functions $\kappa=\kappa(x)$. However, we choose the parameter
$\kappa$ to be spacetime independent for the three concrete cases
discussed in Sec.~\ref{sec:Three-modMax-theories}.

From \eqref{eq:LagMMflat}, the modified Maxwell equation in flat spacetime
is given by
\begin{equation}
\partial_a \left( F^{ab}+\kappa^{abcd} F_{cd}\right) =0.
\end{equation}
The electromagnetic theory \eqref{eq:LagMMflat} allows for maximal photon velocities
different from $c=1$, which clearly indicates the breaking of Lorentz invariance.
See, e.g., Refs. \cite{BaileyKostelecky2004,KosteleckyMewes2002,
KlinkhamerRisse2008,KlinkhamerSchreck2008}
for further details of the simplest version of modified Maxwell theory
in Minkowski spacetime and physical bounds on its 19 spacetime-independent
parameters.

\subsection{Nonbirefringent modified Maxwell theory coupled to gravity}

The vierbein formalism is particularly well-suited for
describing Lorentz-violating theories, since it allows to distinguish
between  local
Lorentz and general coordinate transformations \cite{Kostelecky2004}
and to set the torsion identically to zero. A minimal coupling procedure
then yields the following Lagrange density for the photon part of  the action:
\begin{subequations}\label{eq:LagMMgrav}
\begin{eqnarray}
\mathcal{L}_\text{modMax}
&=&
  -\sqrt{g}\,
  \bra{\frac{1}{4}\,  F_{\mu\nu}F_{\rho\sigma}\,g^{\mu\rho}g^{\nu\sigma}
  +\frac{1}{4}\, \kappa^{\mu\nu\rho\sigma}\,
  F_{\mu\nu}F_{\rho\sigma} }\ ,
\label{eq:LagMMgrav-L}
\\[2mm]
 \kappa^{\mu\nu\rho\sigma}
 &\equiv&
  \kappa^{abcd}\,e^\mu_{\;\;a}\,
  e^\nu_{\;\;b}\,e^\rho_{\;\;c}\,e^\sigma_{\;\;d}\,.
\label{eq:LagMMgrav-kappa-tensor}
\end{eqnarray}
\end{subequations}
Here, and in the following, the numbers $\kappa^{abcd}(x)$
are considered to be fixed parameters, with no field equations of their own.
For the sake of simplicity, the dependence on $x^\mu$ will be suppressed
in the following, $\kappa^{abcd}(x)\equiv \kappa^{abcd}$,
unless stated otherwise.

The purely gravitational part of the Lagrange density is given by the
standard Einstein--Hilbert term~\cite{Wald1984},
\begin{equation}\label{eq:Lgrav}
\mathcal{L}_\text{grav}=\sqrt{g}\,\frac{1}{16\pi}\,R,
\end{equation}
with the Ricci curvature scalar $R$ from the metric $g_{\mu\nu}$.
The complete action is then
\begin{equation}\label{eq:fullaction}
\mathscr{S}=\int d^4x
\left(\mathcal{L}_\text{grav}+\mathcal{L}_\text{modMax} \right),
\end{equation}
where the integration domain still needs to be specified.

\subsection{Field equations}

The variational principle for variations of
the action \eqref{eq:fullaction} with respect to the gauge-field $A_\mu(x)$
results in the following Euler--Lagrange equation
of the photon coupled to gravity:
\begin{equation}\label{eq:gaugefieldequation}
D_\mu \left(F^{\mu\nu}+\kappa^{\mu\nu\alpha\beta}F_{\alpha\beta}\right)=0\;,
\end{equation}
with spacetime-covariant derivative $D_\mu$.
Similarly, variation of the action \eqref{eq:fullaction} with respect
to the vierbein
$\vierbein{\mu}{a}(x)$ leads to the following Einstein equation:
\begin{eqnarray}\label{eq:Einsteinequation}
\hspace*{-5mm}
G^{\mu\nu}&=& 8\pi\, T_\text{vierbein}^{\mu\nu}
\equiv \frac{8\pi}{\sqrt{g}}\;\frac{\delta
  \mathscr{S}_\text{modMax}}{\delta e_{\mu}^{\;\;a}}\;e^{\nu a},
\end{eqnarray}
where $G^{\mu\nu}\equiv R^{\mu\nu}-(1/2)\,R\, g^{\mu\nu}$
denotes the standard Einstein tensor.

Even though the explicit expression for the energy-momentum tensor
is not needed in this article, we give it for completeness:
\begin{subequations}\label{eq:Tmunu+widetildeTmunu}
\begin{eqnarray}
\hspace*{-6mm}
T_\text{vierbein}^{\mu\nu}&\!\!=\!\!&-
\bigg(
F^{\mu\rho}F^{\nu\sigma}\,g_{\rho\sigma}
+\kappa^{\mu\beta\rho\sigma}F^{\nu}_{\;\;\beta}F_{\rho\sigma}
\nonumber\\
\hspace*{-6mm}&\!\!\!\!&
-\frac{1}{4}\;g^{\mu\nu}
\left(F^{\rho\sigma}F_{\rho\sigma}
+\kappa^{\alpha\beta\rho\sigma}F_{\alpha\beta}F_{\rho\sigma}\right)\bigg),
\label{eq:Tmunu}\\
\hspace*{-6mm}
\widetilde{T}_\text{vierbein}^{\mu\nu}&\!\!=\!\!&
- \bigg(\big(1-(\kappa/2)\, \xi^{\rho}\xi_\rho\big)
F^{\mu\rho}F^{\nu\sigma}\,g_{\rho\sigma}
+\kappa\left(F^{\nu}_{\;\;\sigma}F^{\mu}_{\;\;\rho}\, \xi^{\sigma}\xi^{\rho}
+F^{\nu\sigma}F_{\rho\sigma}\xi^{\mu}\, \xi^\rho\right)
\nonumber\\
\hspace*{-6mm}&&
-\frac{1}{4}\; g^{\mu\nu}\Big(\big(1-(\kappa/2)\, \xi^\rho\xi_\rho\big)
      F_{\alpha\beta}F^{\alpha\beta}+
      2\kappa\, F^{\alpha\beta}F_{\alpha\gamma}\, \xi_{\beta}\xi^\gamma
      \Big)\bigg),
\label{eq:widetildeTmunu}
\end{eqnarray}
\end{subequations}
where the first expression holds for the general modified
Maxwell theory and the second (distinguished by a tilde) for the
nonbirefringent sector defined by \eqref{eq:ansatznonbire}--\eqref{eq:ansatzxi}.
Clearly, the relevant expression \eqref{eq:widetildeTmunu}
contains an asymmetric part for generic $\xi^\nu$
with $F_{\mu\nu}\,\xi^\nu \ne 0$.

Since the energy-momentum tensor $T_\text{vierbein}^{\mu\nu}$ of
explicitly Lorentz-violating theories is, in general,
neither symmetric
nor covariantly conserved~\cite{Kostelecky2004}, it is,
\emph{a priori}, not clear that theory \eqref{eq:fullaction} is well
defined or that it has nontrivial solutions at all. Since
it will be sufficient for our purpose to treat the photon as a
test particle described by the modified Maxwell equation
\eqref{eq:gaugefieldequation} in a given spacetime
background, we may safely ignore this issue. However,
it is not difficult to show that theory \eqref{eq:fullaction} has
nontrivial solutions for both the gravitational and photonic fields.

\setcounter{equation}{0}
\section{Effective metric for modified-Maxwell-theory photons}
\label{sec:Effective-metric}

\subsection{Photonic part of the action}

With the particular nonbirefringent \emph{Ansatz}
\eqref{eq:ansatznonbire}--\eqref{eq:ansatzxi},
the photonic Lagrange density \eqref{eq:LagMMgrav} simplifies to
\begin{equation} \label{eq:effectivelagrangian}
\mathcal{L}_\text{modMax} =
-\sqrt{g}\, \left(1-\,\frac{1}{2}\,\kappa(x)\,
  \xi^\rho\xi_\rho\right)\,\frac{1}{4}\,
  F_{\mu\nu}F_{\rho\sigma}\,
  \widetilde{g}^{\mu\rho}\widetilde{g}^{\nu\sigma},
\end{equation}
for $\xi^\rho\xi_\rho = g_{\rho\sigma}\,\xi^\rho\xi^\sigma =\pm 1$
and the following \emph{effective} metric and inverse metric:
\begin{subequations}\label{eq:effectivebackground}
\begin{eqnarray}
\widetilde{g}_{\mu\nu}(x)&=&
g_{\mu\nu}(x)-\frac{\kappa(x)}{1+\kappa(x)\,\xi^\rho\xi_\rho/2}\;
\xi_{\mu}(x)\xi_{\nu}(x)\,,
\label{eq:effectivemetric}
\\[1mm]
\widetilde{g}^{\mu\nu}(x)  &=&
  g^{\mu\nu}(x)+\frac{\kappa(x)}{1-\kappa(x)\,
  \xi^\rho\xi_\rho/2}\;\xi^{\mu}(x)\xi^{\nu}(x)\,,
  \label{eq:effectivemetric-inverse}
\end{eqnarray}
\end{subequations}
which can be verified to satisfy
$\widetilde{g}^{\mu\nu}\widetilde{g}_{\nu\rho}=\delta^{\mu}_{\;\;\rho}$.

As will be demonstrated in the next subsection,
the effective metric is a useful mathematical tool to describe the
propagation of  nonstandard photons obeying the field equation
\eqref{eq:gaugefieldequation} in a given spacetime
background.\footnote{Note that \eqref{eq:effectivelagrangian}
still contains the square root of the determinant of the
original spacetime metric $g_{\mu\nu}$. For the explicit cases to be
discussed in Sec.~\ref{sec:Three-modMax-theories},
it turns out that $\sqrt{g}$ is simply proportional to
$\sqrt{\widetilde{g}}$ and that $\kappa$ is spacetime independent. This then
implies that \eqref{eq:effectivelagrangian} for these special cases equals,
up to an over-all constant, the standard Maxwell action in terms of
the effective metric \eqref{eq:effectivebackground}.}
However, unless explicitly stated otherwise, all lowering or raising of
indices is understood to be performed by contraction with
the original background metric $g_{\mu\nu}$ or its inverse $g^{\mu\nu}$.

\subsection{Dispersion relation and geometrical optics}

The Lagrange density \eqref{eq:effectivelagrangian} is proportional
to a Lagrange density for a
standard photon moving in the effective gravitational field
\eqref{eq:effectivebackground}. We will now demonstrate
that photon trajectories (in the geo\-me\-trical-optics
approximation) are indeed given by null geodesics in this effective metric.

Consider a plane-wave \emph{Ansatz},
\begin{equation}\label{eq:planewaveansatz}
A_\mu (x)=C_\mu(x)\, e^{iS(x)}\,,
\end{equation}
in the Lorentz gauge $D_\mu A^\mu=0$.
As usual, we define the wave vector to be normal to surfaces of equal phase,
\begin{equation}\label{eq:defkmu}
k_\mu\equiv \partial_\mu S\, .
\end{equation}
Inserting \emph{Ansatz} \eqref{eq:planewaveansatz} into the field
equation \eqref{eq:gaugefieldequation} gives
the following dispersion relation:
\begin{subequations}\label{eq:kmucondition}
\begin{equation}\label{eq:kmucondition-kkg}
k_\mu k_\nu\, g^{\mu\nu}=
-\frac{\kappa}{1-\kappa\,\xi^\rho\xi_\rho\,/2}\,
\Big(g^{\mu\nu}\,\xi_{\mu}k_\nu\Big)^2\, ,
\end{equation}
or equivalently
\be\label{eq:kmucondition-kkwideltildeg}
k_\mu k_\nu \,\widetilde{g}^{\mu\nu} =0 \, .
\ee
\end{subequations}
The right-hand side
of \eqref{eq:kmucondition-kkg},
with an over-all factor $\kappa$,
determines the change in the photon dispersion relation due to the
Lorentz-violating part in the Lagrange density \eqref{eq:LagMMgrav}.
The vector
\begin{equation}\label{eq:tangentvector}
\widetilde{{k}}^\mu\equiv\widetilde{g}^{\mu\nu}k_\nu=\dot{x}^\mu
\end{equation}
is tangent to geodesics $x^\mu(\lambda)$ with respect to the effective
metric
$\widetilde{g}_{\mu\nu}$. Here, and in the following,
an overdot denotes differentiation with respect to the
affine parameter $\lambda$.

In order to avoid obvious difficulties with causality,
we intend to restrict our considerations to a subset of
theories without spacelike photon trajectories. As discussed
above, the tangent vector to a photon path
is given by \eqref{eq:tangentvector}. The condition then reads
\begin{equation}
\dot{x}^\mu\, g_{\mu\nu}\, \dot{x}^\nu
=k_\mu\,\widetilde{g}^{\mu\nu}\,g_{\nu\alpha}\,
\widetilde{g}^{\alpha\beta}\,k_\beta\geq
0\,.
\end{equation}
Using definition \eqref{eq:effectivemetric-inverse}, we find
\begin{equation}
\kappa(x)+\frac{1}{2}\,\big(\kappa(x)\big)^2\,\xi^\rho\xi_\rho\geq 0\, ,
\end{equation}
which is satisfied by $\kappa(x)\geq 0$ or $\kappa(x)\leq -2$ for timelike
$\xi^\mu$ and by $0\leq \kappa(x)\leq 2$ for spacelike $\xi^\mu$.
Since we are only interested in small deformations of standard
electrodynamics,
we restrict our considerations to
\begin{equation}\label{eq:kappacausal}
0\leq\kappa(x)\ll 1\,,
\end{equation}
which ensures the causality of the theory for both choices of $\xi^\mu$, at
least as far as the signal-propagation velocity is concerned.

In the above discussion, we have neglected derivatives of the amplitude
$C_\mu(x)$ of \emph{Ansatz} \eqref{eq:planewaveansatz}
and a term involving the Ricci tensor (the typical length scale of $A_\mu$
is assumed to be much smaller than the length scale of the
spacetime background).
See, e.g., Refs.~\cite{Wald1984,BornWolf1999}
for further discussion of the geometrical-optics approximation.

In the next section, three different examples of modified Maxwell theory
in a Schwarz\-schild spacetime background will be presented.
Up to now, $\kappa(x)$ was allowed to be a function of the
coordinates, but for the following three cases
we take $\kappa$ to be a constant, with values in the range
\eqref{eq:kappacausal}.

\setcounter{equation}{0}
\section{Three modified Maxwell theories in a Schwarzschild background}
\label{sec:Three-modMax-theories}

In this section, the spacetime metric is considered to be fixed to that
of a Schwarz\-schild black hole,
\begin{equation}\label{eq:SS-lineelement}
ds^2=\left(1-2 M/r\right) dt^2-\left(1-2 M/r\right)^{-1}
dr^2 -r^2 d\Omega^2\,,
\end{equation}
with parameter $M$ interpreted as the central mass.

The Lorentz-violating parameters $\xi^\mu$ of the following three examples
(\emph{Ca\-ses 1--3}) are introduced by hand and are,
for the moment, not provided by an underlying theory.
But the \emph{Case--1} background field $\xi^\mu$ can be obtained
from the ghost-condensate model of Ref.~\cite{ArkaniHamed-etal2004}
and, in Appendix~\ref{sec:appendix}, we demonstrate that it is also possible
to obtain the \emph{Case--2} background field $\xi^\mu$
as the solution of a dynamic model [\emph{Case--3} is more difficult,
as will be explained in Footnote~\ref{ftn:case3-dynamics} of the appendix].

\subsection{Case 1: Ricci-flat effective metric with changed
  horizon}\label{subsec:Case1}

As a first example, consider nonbirefringent modified-Maxwell-theory photons
moving in the standard Schwarz\-schild background \eqref{eq:SS-lineelement}.
For the parameters $\xi^\mu$ entering \eqref{eq:effectivelagrangian}, we choose
\begin{equation} \label{eq:xi-case-1}
\xi^\mu\,\big|_\text{Case 1}=\bra{\frac{1}{1-2M/r},\, -\sqrt{2M/r},\,
  0,\, 0}
\end{equation}
and take constant $\kappa$ with $0 < \kappa \ll 1$
(the $\kappa=0$ case corresponds to standard Maxwell theory).
Defining two auxiliary parameters
\begin{equation}\label{eq:def-epsilon}
\epsilon\equiv\frac{\kappa}{1-\xi^\rho\xi_\rho\,\kappa/2}
        =     \frac{\kappa}{1-\kappa/2}> 0,
\quad
\chi\equiv \frac{\epsilon}{1+\epsilon} > 0,
\end{equation}
the corresponding effective background \eqref{eq:effectivemetric} reads
\begin{eqnarray}\label{eq:metric-case-1}
\left .d\widetilde{s}^2\right|_\text{Case 1} &=&
\bra{1-2 M/r -\chi} dt^2 - \,2\chi\;\frac{\sqrt{2
    Mr}}{r-2M}\,dt\,dr \nonumber \\
 && \!\!\!\!\!\!- \fr{r}{r-2M} \bra{1 + \chi\; \frac{2 M}{r-2M} } dr^2
 -r^2 d\Omega^2\,. \label{effline}
\end{eqnarray}
From the effective metric \eqref{eq:metric-case-1} at $r \gg 2\,M$,
the maximal photon velocity
is found to be given by $v_{\gamma,\,\text{max}}=\sqrt{1-\chi}$,
which, for small but positive $\chi$, is less than the maximal velocity
$c=1$ of Lorentz-invariant matter, according to \eqref{eq:SS-lineelement}.

An appropriate coordinate transformation,
\begin{equation}
dt = \sqrt{1+\epsilon}\;dT + \chi\;\frac{\sqrt{2 Mr}}{r-2M}\;
     \frac{1}{1-2 M/r-\chi}\;dr\,,
\end{equation}
reveals that \eqref{eq:metric-case-1} is just a standard Schwarz\-schild
background,
but with a rescaled mass:
\begin{subequations}\label{eq:SSmodmassexplicit}
\begin{eqnarray}\label{eq:SSmodmass-linelement}
d\widetilde s^2\,\big|_\text{Case 1}
&=&
\bra{1-2 \widetilde{M}/r} dT^2-\bra{1-2 \widetilde{M}/r}^{-1}
dr^2 -r^2 d\Omega^2\,,
\\[1mm]
\widetilde{M} &\equiv& M\bra{1+\epsilon}.
\label{eq:Case1-widetildeM}
\end{eqnarray}
\end{subequations}
Therefore, it is clear that the event horizon of these modified-Maxwell-theory
photons is different from the standard Schwarz\-schild event horizon at
$r=r_\text{Schw}\equiv 2M$.\footnote{The
effective background \eqref{eq:metric-case-1}
transformed to the Lema\^{i}tre reference frame agrees with the effective
metric obtained in Ref.~\cite{DubovskySibiryakov2006} for a minimally
coupled scalar field interacting with the ghost condensate.}
Specifically, these nonstandard
photons would have a horizon at $r=2\widetilde{M}=2M\bra{1+\epsilon}$,
which lies outside the horizon of standard Lorentz-invariant matter
(protons, neutrons, and electrons) at $r=2M$. A theory consisting of
these Lorentz-violating photons and standard Lorentz-invariant matter
(protons, neutrons, and electrons) would suffer the same difficulties
concerning the generalized second law of thermodynamics as described in
Refs.~\cite{DubovskySibiryakov2006,Eling-etal2007,JacobsonWall2008}.

At the level of a \emph{Gedankenexperiment},
let us try to give a concrete realization of the very simple entropy-reducing
process discussed in Ref.~\cite{JacobsonWall2008}.
The theory we consider is modified quantum electrodynamics (QED) with
an action containing, first,
the standard Dirac terms~\cite{JauchRohrlich1976,BirrellDavies1982}
for protons, neutrons, and electrons,\footnote{These standard-model fermions
may also have standard Lorentz-invariant strong interactions so that nuclei
can be formed, which are needed for the walls of the box used in
the \emph{Gedankenexperiment} and the rope attached to the box.
Otherwise, the box walls have to be made of frozen hydrogen
and a frozen-hydrogen chain needs to be fabricated (in order to replace the rope),
all within the possibilities of a \emph{Gedankenexperiment}
operating at $T < 14\;\text{K}$ and normal pressure.}
and, second, the \emph{Case--1} modified-Maxwell-theory term,
explicitly given by \eqref{eq:effectivelagrangian}, \eqref{eq:effectivebackground},
and \eqref{eq:xi-case-1},
for constant $\kappa$ at a fixed nonzero value $0 < \kappa \ll 1$.
Calling these modified-Maxwell-theory photons ``A--type particles'' and
these standard fermions (protons, neutrons, and electrons) ``B--type particles,''
we can just quote the authors of Ref.~\cite{JacobsonWall2008}:
``To violate the heat version of the second law, one can suspend a B--box
containing thermal A--radiation on a B--rope, and lower the box past
the A--horizon. The A--heat can then be discarded in the black hole, thus
converting it into work at infinity, without changing the exterior
appearance of the black hole nor using up any exterior fuel.''
Two crucial ingredients here are, first, the existence of A--B
interactions and, second, the possibility that certain incoming positive-energy
A--photons fall into the negative-energy states
of the ``ergoregion'' $r \in (2M,\,2\widetilde{M})$, which can be
described by the effective Schwarzschild metric \eqref{eq:SSmodmassexplicit}
transformed to regular coordinates such as Kruskal coordinates~\cite{Wald1984}.
More specifically,
it is possible to repeat the detailed calculation of Ref.~\cite{UnruhWald1983}
to show that the final increase of the black-hole entropy is strictly less
than the initial entropy of the A--radiation dropped in.
In this way, the generalized second law of thermodynamics would be violated
for the modified QED theory with \emph{Case--1} modified-Maxwell-theory
photons (A--type particles) and standard Lorentz-invariant fermions
(B--type particles).

\subsection{Case 2: Non-Ricci-flat effective metric with unchanged horizon}
\label{subsec:Case2}

Again, consider the Schwarz\-schild spacetime background
\eqref{eq:SS-lineelement} in the
coordinate patch $r>2M$, but now take
\begin{equation} \label{eq:xi-case-2}
\xi^\mu\,\big|_\text{Case 2}=\left(0,\, \sqrt{1-2 M/r},\, 0,\, 0\right)
\end{equation}
and constant $\kappa$ with $0 < \kappa \ll 1$.
The effective line element \eqref{eq:effectivemetric} for the photon
field is then given by
\begin{equation}\label{eq:metric-case-2}
\left . d\widetilde s^2\right|_\text{Case 2}=
\bra{1-2 M/r} dt^2-\frac{1}{1-\eta} \bra{1-2 M/r}^{-1}
dr^2 -r^2 d\Omega^2,
\end{equation}
in terms of the auxiliary Lorentz-violating parameter
\begin{equation}\label{eq:def-eta}
\eta \equiv \frac{\kappa}{1-\xi^\rho\xi_\rho\,\kappa/2}
     =      \frac{\kappa}{1+\kappa/2} > 0\,.
\end{equation}
From the effective metric \eqref{eq:metric-case-2} at $r \gg 2\,M$,
the maximal radial photon velocity
is found to be given by $v_{\gamma,\,\text{rad,\,max}}=\sqrt{1-\eta}<1$,
for small but positive $\eta$.
The maximal tangential photon velocity is not affected
by the Lorentz violation, $v_{\gamma,\,\text{tang,\,max}}=1$.

Although the resulting effective metric \eqref{eq:metric-case-2} still has
a coordinate singularity at $r=2M$,
it is no longer a Schwarz\-schild solution:
the effective Ricci scalar does not vanish and is, in fact, given by
\begin{equation}
R\big[\,\widetilde{g}\,\big]=-2 \eta/r^2\,.
\end{equation}
However, the temperature (derived from the periodicity of the Euclidean time
variable $\tau \equiv \mathrm{i}\,t$) is given by $T=(8\pi M)^{-1}$, which
equals the Hawking temperature~\cite{Hawking1975}
of a Schwarzschild black-hole of mass
$M$ in standard general relativity, $T_\text{H}=(8\pi M)^{-1}$. That is,
the effective background for the photons appears to have
the same temperature as the
standard Schwarz\-schild black hole, irrespective of the Lorentz-violating
parameter $|\eta| < 1$. For this particular Lorentz-violating
theory,\footnote{In Sec.~\ref{subsec:Case2-outbound-photon-},
we will also show by explicit calculation that the expected photon horizon at
$r=2M/(1-\eta)>2M$ does not occur.} there is no obvious conflict with the
generalized second law of thermodynamics and it is not possible to construct
a perpetuum mobile of the second kind, at least along the lines suggested by
Refs.~\cite{DubovskySibiryakov2006,Eling-etal2007}.

In order to get some insight into the possible meaning of the
non-Ricci-flat effective metric \eqref{eq:metric-case-2},
let us seek a perfect-fluid matter source which would give
rise to this metric by the Einstein equation
$\widetilde{G}^\alpha_{~\beta}\equiv G^\alpha_{~\beta}\big[\,\widetilde{g}\,\big]
=8\pi\,T^\alpha_{~\beta}$.
The energy-momentum tensor is
$T^\alpha_{~\beta}=(\rho+P)u^\alpha u_\beta - P\,g^\alpha_{~\beta}$
and the comoving four-velocity of the fluid is assumed to be
$u^\alpha=\left((1-2M/r)^{-1/2},0,0,0\right)$. Since
the Einstein tensor corresponding to the metric \eqref{eq:metric-case-2} is
$\widetilde{G}^\alpha_{~\beta}=\text{diag}\big(\eta/r^2,$ $\eta/r^2,0,0\big)$,
we obtain the energy density $8\pi \rho=\widetilde{G}^t_{~t}= \eta/r^2$ and
the isotropic pressure $24\pi P=-\widetilde{G}^r_{~r}= -\eta/r^2$.
Interestingly, the required perfect fluid corresponds
to some type of modulated ``dark energy,''
with constant equation-of-state parameter $w \equiv P/\rho =-1/3\,$.

\subsection{Case 3:  Ricci-flat effective metric with unchanged
  horizon}\label{subsec:Case3}

It is even possible to have a situation where the effective metric is
again the
Schwarz\-schild metric \eqref{eq:SS-lineelement}
with the same mass $M$, just as if the Lorentz violation were not
present
(but it is and gives rise to measurable effects, as will be
shown in Sec.~\ref{subsec:Bending-of-light}).
This happens, for instance, if the Lorentz-violating tensor
$\widetilde{\kappa}_{\mu\nu}$ is spatially isotropic
in the coordinate patch $r>2M$:
\begin{equation} \label{eq:xi-case-3}
\xi^\mu\,\big|_\text{Case 3}=\left(\bra{1-2 M/r}^{-1/2},0,0,0\right),
\end{equation}
taking, again, constant $\kappa$ with $0 < \kappa \ll 1$.
Now, the effective line element \eqref{eq:effectivemetric} is
\begin{equation} \label{eq:metric-case-3}
\left. d\widetilde s^2\right|_\text{Case 3}=\frac{1}{1+\epsilon}
\bra{1-2 M/r} dt^2-\bra{1-2 M/r}^{-1} dr^2 -r^2
d\Omega^2,
\end{equation}
with $\epsilon$ defined by \eqref{eq:def-epsilon}.
From the effective metric \eqref{eq:metric-case-3} at $r \gg 2\,M$,
the maximal photon velocity
is found to be given by $v_{\gamma,\,\text{max}}=1/\sqrt{1+\epsilon}<1$,
for small but positive $\epsilon$.
The effective metric \eqref{eq:metric-case-3}
becomes, of course, the original Schwarz\-schild metric by rescaling
$t=\sqrt{1+\epsilon}\;\widetilde t$.

Observe that this equivalence of metrics only means that light rays in
the effective metric follow the same paths as standard photons in the
original Schwarz\-schild metric.
But, in the observer's reference frame, these light rays travel with a
velocity different from $c$ since the dispersion relation is still given by
\eqref{eq:kmucondition}. In fact, consider a stationary observer with
four-velocity $u^\mu =\left((1-2M/r)^{-1/2},\, 0,\, 0,\, 0\right)$, in whose
local inertial coordinate system the vierbein has components
\begin{equation} \label{eq:vierbein-observer}
\vierbein{t}{0} = \sqrt{1-2M/r} =  \bra{\vierbein{r}{1}}^{-1} ,
\quad  \vierbein{\theta}{2} = r, \quad   \vierbein{\phi}{3}= r\sin\theta\,,
\end{equation}
with the other components vanishing.
The observer measures the frequency $\omega = g_{\mu\nu}\,k^\mu u^\nu = k^0$,
where the wave vector $k^\mu$ satisfies \eqref{eq:kmucondition}.
One readily obtains the quadratic dispersion relation
\begin{equation} \label{eq:dispersion}
\omega({\bf k})^2
= \frac{1-\kappa/2}{1+\kappa/2}\;|{\bf k}|^2
= \frac{1}{1+\epsilon}\;|{\bf k}|^2\, ,
\end{equation}
for wave vector ${\bf k}\equivdef(k^1,k^2,k^3)$.
With $ \kappa>0$, it is then clear that light rays travel with maximal velocity
\begin{equation}\label{eq:vmax}
v_{\gamma,\,\text{max}} = \sqrt{\frac{1-\kappa/2}{1+\kappa/2}} < 1\,,
\end{equation}
which agrees with the previous value given a few lines
below \eqref{eq:metric-case-3}. Of course, the dispersion relation is more
complicated than \eqref{eq:dispersion} for nonisotropic cases, e.g., the
choices \eqref{eq:xi-case-1} and \eqref{eq:xi-case-2} for the parameters
$\xi^\mu$.

As will be demonstrated in Sec.~\ref{sec:Redshift-and-bending},
photons traveling in an
effective background described by \eqref{eq:metric-case-3} have measurable
properties different from standard photons moving in a Schwarz\-schild background,
even though the background \eqref{eq:metric-case-3} is related to a standard
Schwarzschild background by a mere coordinate transformation. This
reflects the fact that theory \eqref{eq:LagMMgrav} is not invariant
under general coordinate transformations.
This is especially relevant if the modified-Maxwell-theory photon
is also coupled to standard Lorentz-invariant matter.
An observer with standard measuring devices still perceives a standard
Schwarzschild background described by line element \eqref{eq:SS-lineelement}.
It is the presence of the two metrics \eqref{eq:SS-lineelement} and
\eqref{eq:metric-case-3}, which cannot be brought to a standard
Schwarzschild background simultaneously, that gives rise to the
measurable properties derived in the next section.

\setcounter{equation}{0}
\section{Gravitational redshift and bending of light}
\label{sec:Redshift-and-bending}

In this section, we will calculate the gravitational redshift
and the bending of light for the three modified Maxwell theories
of Sec.~\ref{sec:Three-modMax-theories}, there
called \emph{Cases 1--3}.
Furthermore, we will explicitly demonstrate that the effective
metric \eqref{eq:metric-case-2} of \emph{Case 2} does not have a
photon horizon at the naively expected value $r=2M/(1-\eta)$.
We start, however, with a brief discussion of the horizons
present in these theories.

\subsection{Horizons}
\label{subsec:horizons}

Just as for the standard Schwarzschild geometry,
the coordinate $t$ of the \emph{Case--1} line element \eqref{eq:metric-case-1}
becomes spacelike and $r$ becomes timelike for $r<2\widetilde{M}$.
A photon emitted from position $r<2\widetilde{M}$ must always travel forward
in
``time,'' i.e., to decreasing $r$ and can, therefore, never leave the
region $r<2\widetilde{M}$.
This picture is confirmed by studying the geodesic paths of photons
for the inside region $r<2\widetilde{M}$. Hence, $r=2\widetilde{M}$ is a
genuine event horizon
for \emph{Case 1}.

A similar argument holds for \emph{Cases 2--3} if, on the one hand,
$\widetilde{M}$ is replaced by $M$ and if, on the other hand,
the following assumptions for the parameters
$\xi^\mu$ in the interior region $r<2M$ are made:
\begin{subequations} \label{eq:xi-inside-horizon}
\begin{eqnarray} \label{eq:xi2-inside-horizon}
\xi^\mu\,\big|_\text{Case 2, interior}&=&
\left(0,\, \sqrt{2 M/r-1},\, 0,\,0\right),
\\[1mm]
 \label{eq:xi3-inside-horizon}
\xi^\mu\,\big|_\text{Case 3, interior}&=&
\left(1/\sqrt{2 M/r-1},0,0,0\right).
\end{eqnarray}
\end{subequations}
With these additional assumptions, the $r=2M$ surface is an event horizon
for \emph{Case 2} and \emph{Case 3}.
Independently of these additional assumptions, we will show in
the next subsection that this $r=2M$ surface
is an infinite-redshift horizon for a distant observer.

Adding standard Lorentz-invariant matter to the modified-Maxwell-theory photons,
there is then an outer horizon for
the \emph{Case--1} photons and an inner one for the matter particles
(the ergoregion in between these horizons is crucial for the apparent
violation of the generalized second law of thermodynamics,
as discussed in the last paragraph of Sec.~\ref{subsec:Case1}).
The horizons for \emph{Case--2} and \emph{Case--3}
modified-Maxwell-theory photons
coincide with the horizon for standard matter.

\subsection{Redshift}
\label{subsec:Redshift}

To discuss the redshift, we solve explicitly for a light ray
described by a tangent vector $k_\mu$, which obeys
\begin{equation}
k_\mu \, \widetilde{g}^{\mu\nu}\, k_\nu=0,\quad
\widetilde{g}^{\mu\nu}\, k_\nu\, \widetilde{D}_\mu k_\lambda=0,
\end{equation}
where $\widetilde{D}_\mu$ denotes covariant differentiation with respect
to the effective metric $\widetilde{g}_{\mu\nu}$.
We assume that the hypothetical measuring device is built of
standard matter (protons, neutrons, and electrons), so that no additional
Lorentz violation is introduced by the measuring process,
at least to leading order in $\kappa$.
The frequency measured by an observer at spacetime point $P_i$ with
arbitrary four-velocity $u^\mu_i$  is then given by
\begin{equation}
\omega_{P_i}= k_\mu\,  g^{\mu\nu}\,  u_\mu\, \Big|_{P_i}\, .
\end{equation}
In the following, we restrict the discussion
to a stationary observer with $u^\mu=t^\mu/(t^\nu t_\nu)$,
where $t^\mu$ is the timelike Killing field of the observer
Schwarz\-schild background.

A straightforward calculation
for all three cases of Sec.~\ref{sec:Three-modMax-theories}
then shows that the redshift equals that of standard photons:
\begin{equation}\label{eq:redshift}
\frac{\omega_1}{\omega_2}\,\Bigg|_\text{Case\,1,2,3}=
\sqrt{\frac{1-2G_\text{N} M/(r_2\,c^2)}{1-2G_\text{N} M/(r_1\,c^2)}}\,,
\end{equation}
where $G_\text{N}$ and $c$ have been restored temporarily.
However, \eqref{eq:redshift} is only valid for $P_i$ lying outside the
corresponding photon horizon, i.e.,
$r_i> 2\widetilde{M}$  for \emph{Case 1} and
$r_i> 2M$ for \emph{Case 2} and \emph{Case 3}.

Now, consider the following \emph{Gedankenexperiment}. A massive gamma-ray
source is falling towards the Schwarzschild singularity
and emits photons isotropically (in its rest frame).
An observer near infinity, $r_2\gg 2M$, measures then the following
redshift:
\begin{equation}\label{eq:observed-redshift}
z \approx 1/\sqrt{1-2M/r_1}-1,
\end{equation}
where $r_1$ is the emission point of the gamma-ray photon.
(We neglect possible additional Doppler-shift effects, in order to
simplify the
discussion.)
For the \emph{Cases 2--3}, the redshift \eqref{eq:observed-redshift}
goes to infinity when
the source approaches $r_1=2M$. For \emph{Case 1}, however, the maximal
redshift an observer at infinity will find is
\begin{equation}
z=1/\sqrt{1-2M/\big(2M(1+\epsilon)\big)}-1
=\sqrt{(1+\epsilon)/\epsilon}-1 ,
\end{equation}
which is large but finite for small but positive $\epsilon$.

\subsection{Bending of light}\label{subsec:Bending-of-light}

It is well known that standard light rays propagating in the exterior
region
of the Schwarz\-schild spacetime experience deflection by the central
mass. We now analyze this deflection process for light rays obeying the
modified Maxwell equation \eqref{eq:gaugefieldequation}.

For all the effective metrics of Sec.~\ref{sec:Three-modMax-theories},
we still find a timelike and a rotational Killing field, given by
$t^\mu=\left(\partial/\partial t\right)^\mu$ and
$\psi^\mu=\left(\partial/\partial \phi\right)^\mu$, respectively. Also,
just as for the usual Schwarz\-schild metric,
suitable rotations of the coordinate system are able to confine
a geodesic to the equatorial plane $\theta=\pi/2$.

For \emph{Case 2} and \emph{Case 3}, the
effective metric is still diagonal. For a null geodesic
$x^{\mu}(\lambda)$ in these backgrounds, it is then possible to identify
the following constants of motion~\cite{Wald1984}:
\begin{subequations}\label{eq:constants-EL}
\begin{eqnarray}
E &=&
\widetilde{g}_{\mu\nu}\, t^\mu \widetilde{k}^\nu
=\widetilde{g}_{tt}\, \dot{t}\;,
\label{eq:constant-E}
\\[1mm]
L &=& -\widetilde{g}_{\mu\nu}\, \psi^\mu \widetilde{k}^\nu=r^2 \dot{\phi}\;,
\label{eq:constant-L}
\end{eqnarray}
\end{subequations}
where $\dot{x}^\mu(\lambda)= \widetilde{k}^\mu$ denotes the tangent vector.
Making use of these constants of motion, the geodesic equation in these
effective backgrounds reduces to
\begin{equation}
0=\frac{E^2}{\widetilde{g}_{tt}\, \widetilde{g}_{rr}}+\dot{r}^2
-\frac{L^2}{r^2\,  \widetilde{g}_{rr}}\,.
\end{equation}
The equation for the spatial orbit of the light ray is then given by
\begin{equation}
\frac{d\phi}{dr}=\frac{L}{r^2}\,
\frac{1}{\sqrt{-E^2/(\widetilde{g}_{rr}\, \widetilde{g}_{tt})+
    L^2/(r^2\, \widetilde{g}_{rr})}}\, .
\end{equation}
The deflection angle $\delta \phi$  is found to be
\begin{equation}\label{eq:intdeflection}
\delta\phi\equiv-\pi+2 \int_{r_0}^\infty dr\frac{d \phi}{dr}\,,
\end{equation}
where the distance of closest approach, $r_0$, is given by
\begin{equation}
\left.\frac{dr}{d\phi}\right|_{r=r_0}=0\,.
\end{equation}

Inserting the corresponding values for \emph{Case 2}
and using definition \eqref{eq:def-eta} for $\eta$, one finds
\begin{equation}
\left.\frac{d\phi}{dr}\right|_\text{Case 2}=
\frac{L}{r^2\sqrt{1-\eta}}\;\frac{1}{\sqrt{{E^2}-L^2\left(1-2M/r\right)/r^2}}\,,
\end{equation}
which is $1/\sqrt{1-\eta}$ times the standard expression. The total
deflection angle is, therefore, changed
compared to what would be expected for standard Lorentz-invariant photons
by the same factor,
\begin{equation}
\left.\delta \phi\right|_\text{Case 2}
=1/\sqrt{1-\eta}\;\frac{4G_\text{N} M E}{L c^2}
=1/\sqrt{1-\eta}\,\left.\delta \phi\right|_\text{standard}\,,
\end{equation}
where $G_\text{N}$ and $c$ have been restored temporarily.
Recall that $E$ and $L$ have the dimension of length and length square, respectively,
which is consistent with the standard definition~\cite{Wald1984}
of the apparent impact parameter, $b \equiv L/E$.

Similarly, one finds for \emph{Case 3}:
\begin{equation}
\left.\frac{d\phi}{dr}\right|_\text{Case 3}
=
\frac{L}{r^2}\; \frac{1}{\sqrt{{E^2(1+\epsilon)}-L^2\left(1-2M/r\right)/r^2}}\,,
\end{equation}
with $\epsilon$ defined by \eqref{eq:def-epsilon}.
The corresponding integral \eqref{eq:intdeflection} can be evaluated
to first order in $M$. We then obtain for the deflection angle
\begin{equation}
\left.\delta \phi\right|_\text{Case 3}\approx
\sqrt{1+\epsilon}\;\left.\delta \phi\right|_\text{standard}\,,
\end{equation}
with terms of order $\big(G_\text{N} M E/(L c^2)\big)^2$ neglected.

\emph{Case 1} is more subtle. Of course, the calculation in
background \eqref{eq:SSmodmassexplicit} yields
$\delta\phi=4 G_\text{N} \widetilde{M}\widetilde{ E}/(c^2 \widetilde{L})$.
But, now, $\widetilde{E}$ and $\widetilde{L}$ are the constants
of motion in the background \eqref{eq:SSmodmassexplicit}, which still have to be
related to the physically measurable quantities $E_\text{phys}$ and
$L_\text{phys}$, which a stationary observer
would measure at infinity (the observer being stationary with respect
to the original Schwarz\-schild background). The required relation is given by
\begin{equation}
E_\text{phys}=\widetilde{E}/\sqrt{1+\epsilon},\quad
L_\text{phys}=\widetilde{L}\,,
\end{equation}
so that
\begin{equation}
\left.\delta\phi\right|_\text{Case 1}
\approx
\frac{(1+\epsilon)^{3/2}\;4 G_\text{N} M E_\text{phys}}{c^2  L_\text{phys}}
=
(1+\epsilon)^{3/2}\; \delta\phi_\text{standard}\,.
\end{equation}
Here, we have, again, neglected terms of
order $\big(G_\text{N} M E_\text{phys}/(L_\text{phys} c^2)\big)^2$.

For all three cases considered, the deflection of the particular
modified-Maxwell-theory photons is different from that of a standard
Lorentz-invariant photon.\footnote{As there is only one type of photon
in nature, it may be more appropriate to compare, in a \emph{Gedankenexperiment},
the nonstandard deflection of a possible modified-Maxwell-theory photon
and the standard deflection of a Lorentz-invariant proton
at ultrarelativistic energy.}
Still, the impact of the Lorentz violation on the issue of
black-hole thermodynamics depends on the case discussed,
with \emph{Case 1} leading to an apparent contradiction
with the generalized
second law of thermodynamics (cf. the discussion at the end
of Sec.~\ref{subsec:Case1}) but \emph{Case 2} and \emph{Case 3} not.

\subsection{Case 2: Outbound photon}
\label{subsec:Case2-outbound-photon-}

For the effective metric \eqref{eq:metric-case-1} of \emph{Case 1},
a maximal speed of light
$v_{\gamma,\,\text{max}}=c/\sqrt{1+\epsilon}$ is associated with a
modified horizon
$r=\widetilde{r}_\text{Schw}\equiv (1+\epsilon)\, r_\text{Schw}=(1+\epsilon)\, 2M$;
see also Footnote~\ref{ftn:naive-horizon}. It is tempting
to generalize this statement, i.e., to reason that a modified maximal speed
of photons \emph{always}
corresponds to a modified Schwarz\-schild horizon. By the same
reasoning, also a nontrivial modification of the horizon for the effective metric
\eqref{eq:metric-case-2} of \emph{Case 2} would be expected.

The maximal velocity is, however, direction
dependent. For particles moving  ``parallel'' to the parameter
four-vector $\xi^\mu$ of \emph{Case 2},
the maximal achievable velocity is $\sqrt{1-\eta}$. Nontrivial effects
could then be expected for $r=2M/(1-\eta)$, analogous to those
in Refs.~\cite{DubovskySibiryakov2006,Eling-etal2007}.
But it will be shown in the following that a photon which starts
from $r=2M/(1-\eta/2)<2M/(1-\eta)$ can escape,
on a geodesic path, to arbitrary large coordinate $r$ in a finite time $t$.
Hence, $r=2M/(1-\eta)$ is \emph{not} a horizon for the particular type of
photons considered and the intuitive reasoning proves incorrect.
The Schwarz\-schild horizon at $r_\text{Schw}=2M$
persists, manifesting itself through the vanishing of the timelike Killing
field at $r=r_\text{Schw}\equiv 2M$.
Both Lorentz-invariant matter (protons, neutrons, and electrons)
and our particular Lorentz-violating photons
would still have an event horizon at $r=r_\text{Schw}$.

The tangent vector for a radial outgoing geodesic $\dot{x}^\mu(\lambda)$
with dimensionless affine parameter $\lambda$ is given by
the following differential equation:
\begin{equation}
\dot{x}^\mu=E\left(\frac{r}{r-2M},\, \sqrt{1-\eta},\, 0,\, 0\right),
\end{equation}
for $r\geq 2M/(1-\eta/2)$.
Here, $E$ is again the energy-like constant of motion and
the Lorentz-violating parameter $\eta$ is assumed to be small but nonzero,
$0< \eta \ll 1$.
For a geodesic starting at $t=0$ from position
\begin{subequations}\label{eq:Case2-explicit-path-out}
\begin{equation}
r_\text{start}\equiv r_\text{Schw}/(1-\eta/2)=2M/(1-\eta/2),
\end{equation}
the integrated path is explicitly given by
\begin{eqnarray}
x^\mu(\lambda)&=&\left(E\,\tau(\lambda),
E\,\lambda\,\sqrt{1-\eta}+r_\text{start},0,0\right),
\\[1mm]
\tau(\lambda)&=&\lambda+\frac{2M}
{E\sqrt{1-\eta}}\;\ln\left( 1+
\frac{E\,\lambda\,\sqrt{1-\eta}}{r_\text{start}-r_\text{Schw}}\,
\right),
\end{eqnarray}
\end{subequations}
for parameter $\lambda \in [0,\infty)$.
Clearly, arbitrary large positions $r$ can be reached in a finite time $t$
and there is no event horizon at $r=2 M/(1-\eta)$.

\setcounter{equation}{0}
\section{Conclusion}

In this article, we studied the propagation of nonstandard photons
in a given gravitational background and introduced a general method
to describe the geometrical-optics approximation
of nonbirefringent modified Maxwell theory in a given curved-spacetime
background. With  the  nonbirefringent \emph{Ansatz}
\eqref{eq:ansatznonbire}--\eqref{eq:ansatzxi}
for the parameters $\kappa^{\mu\nu\rho\sigma}$ in \eqref{eq:LagMMgrav},
one can, in fact,
describe all phenomena of the geometrical-optics approximation by making
use of an effective  metric $\widetilde{g}_{\mu\nu}(x)$,
where the photons follow null geodesics of this effective metric.
Standard Lorentz-invariant particles (e.g., protons, neutrons, and
electrons) still propagate according to the usual equations of motion
with the original spacetime metric $g_{\mu\nu}(x)$.

Several choices of the parameters $\kappa^{\mu\nu\rho\sigma}$  have
been presented in Sec.~\ref{sec:Three-modMax-theories},
for example, a choice which rescales
the mass value for the effective Schwarz\-schild  metric
(\emph{Case~1}) or a choice which generates an
effective  metric with negative Ricci scalar (\emph{Case~2}).
We find that, in general, a maximal photon propagation velocity smaller
than the causal
velocity $c$ influences the bending of light by the Schwarz\-schild mass
(Sec.~\ref{subsec:Bending-of-light}), but
has not necessarily an effect on the notion of the event horizon.

Lorentz-violating theories with effective metrics that modify the
event horizon (\emph{Case~1} of Sec.~\ref{sec:Three-modMax-theories})
appear to allow for the construction of a perpetuum mobile
of the second kind~\cite{DubovskySibiryakov2006,Eling-etal2007},
bringing them in conflict with the generalized second law of
thermodynamics~\cite{Bekenstein1974,Hawking1975,UnruhWald1983,FrolovPage1993}.
Other Lorentz-violating theories such as those of
\emph{Cases~2--3} of Sec.~\ref{sec:Three-modMax-theories}
avoid such difficulties and
are examples of  theories which incorporate Lorentz violation
but do not appear to violate the generalized second law of thermodynamics.
Hence, we have shown that not every Lorentz-violating theory with
multiple maximal
velocities is in apparent conflict with black-hole thermodynamics.

\section*{Acknowledgments}
We thank M. Schreck for helpful discussions and T. Jacobson for useful comments
on the first version of this article.

\begin{appendix}
\section{ Dynamic model for Case--2 Lorentz Violation}\label{sec:appendix}
\renewcommand{\theequation}{A.\arabic{equation}}\setcounter{equation}{0}

The effective metrics discussed in Sec.~\ref{sec:Three-modMax-theories}
have not been obtained
from an underlying theory but were introduced by hand to show that
not every Lorentz-violating dispersion relation with maximal velocities
different from the causal velocity $c$ leads to modifications of the
black-hole horizon and corresponding difficulties with black-hole thermodynamics.
In this appendix, we present a toy model which yields one of the
proposed effective metrics as a solution of the field equations.
The particular toy model presented here is directly inspired by the
model studied in Ref.~\cite{ArkaniHamed-etal2004}
(related models have been studied in
Refs.~\cite{ArmendarizPiconMukhanovSteinhardt2000,{BabichevMukhanovVikman2006}}
and other references therein).

Consider a gravitating scalar field $\phi$ described by the following action:
\begin{subequations}\label{eq:actioncase2}
\begin{eqnarray}
S&=&\int d^4 x\left(\mathcal{L}_\text{grav}+\sqrt{g} \; m^4\;V\right)\,,
\label{eq:actioncase2-S}\\
V&=&(X+1)^2,
\label{eq:actioncase2-V}\\
X &\equiv&  g^{\mu\nu}\,(\partial_\mu \phi) \, (\partial_\nu \phi)
\label{eq:actioncase2-X}\,,
\end{eqnarray}
\end{subequations}
with the pure-gravity Lagrange density $\mathcal{L}_\text{grav}$
given by \eqref{eq:Lgrav}
and a noncanonical mass dimension $-1$ for the scalar field $\phi$,
so that $X$ is dimensionless.

Two remarks are in order.
First, observe that, with definition \eqref{eq:actioncase2-V}, the sign of
the quadratic kinetic term of the $\phi$ field in \eqref{eq:actioncase2-S}
is standard (non-ghostlike), contrary to the case
of the model of Ref.~\cite{ArkaniHamed-etal2004}.
Second, the explicit potential $V(X)$ in \eqref{eq:actioncase2-V}
can be generalized, provided that there remains a minimum at $X=-1$ with
\begin{equation}\label{eq:potentialconditions}
V(-1)=V^\prime(-1)=0\,,\quad V^{\prime\prime}(-1)> 0\,,
\end{equation}
where the prime denotes differentiation with respect to $X$.

The field equations from \eqref{eq:actioncase2} are
\begin{subequations}\label{eq:fieldequationsapp}
\begin{equation}
G^{\mu\nu}=-4\pi m^4\left(
2 V^\prime\;\partial^\mu\phi\;\partial^\nu\phi + V g^{\mu\nu}\right)\,,
\end{equation}
\begin{equation}\label{eq:fieldequationphi}
D_\mu  \left(V^\prime\,\partial^\mu\phi\right)=0\,,
\end{equation}
\end{subequations}
where $G^{\mu\nu}$ denotes the Einstein tensor defined
under \eqref{eq:Einsteinequation} and $D_\mu$ the covariant derivative.

The flat spacetime solution of the field equations \eqref{eq:fieldequationsapp} has
\begin{subequations}\label{eq:flat-solution}
\begin{eqnarray}
g_{\mu\nu}  &=& \eta_{\mu\nu}\equiv \text{diag}( 1,\, -1,\, -1,\, -1)\,,\\
\phi        &=& n_\mu\,x^\mu\,,
\end{eqnarray}
\end{subequations}
in terms of Cartesian coordinates and a constant spacelike vector $n_\mu$,
that is, $n_\mu n_\nu\, \eta^{\mu\nu}$ $=$ $-1$.
This vector $n_\mu$ may be the source of observable anisotropy (Lorentz
violation) if the scalar field $\phi$ is coupled to standard matter;
cf. Ref.~\cite{KlinkhamerRisse2008} for experimental bounds
in the photon sector.

The Schwarzschild metric and any $\phi$ configuration
with $X\equiv \partial_\mu \phi\, \partial^\mu \phi=-1$
also solve the field equations \eqref{eq:fieldequationsapp}.
Together with the Schwarzschild metric \eqref{eq:SS-lineelement}
in standard spherical coordinates, an explicit solution for the scalar
field $\phi$ over the coordinate patch $r>2M$  is given by
\begin{subequations}\label{eq:Schwarzschild-solution}
\begin{eqnarray}
\overline{g}_{\mu\nu}  &=&
\text{diag}\left(
\left[1-2 M/r\right],\,
-\left[1-2 M/r\right]^{-1},\,
-r^2 ,\,
-r^2 \,\sin^2\theta
\right)\,,\\[2mm]
\overline{\phi} &=&
2 M \,\ln \left(\frac{\sqrt{r}+\sqrt{r-2 M}}{\sqrt{M}}\right) +\sqrt{r\,(r-2M)}\,,
\label{eq:overline-phi}
\end{eqnarray}
\end{subequations}
where $2M$ multiplied by $G_\text{N}/c^2$ is the length parameter of the solution.
The field configuration \eqref{eq:overline-phi} solves the field
equation \eqref{eq:fieldequationphi}
outside the horizon and its gradient is identical to the
\emph{Case--2} background field \eqref{eq:xi-case-2},
\begin{equation} \label{eq:xi-case-2-dynamic}
\xi_\mu\,\big|_\text{Case 2}=  \partial_\mu \overline{\phi}\,,
\end{equation}
with nonvanishing radial component $\partial_r \overline{\phi}=1/\sqrt{1-2M/r}$.

Note that throughout most of the present article (the only exception occurring in
Sec.~\ref{subsec:horizons}), we do not make any assumptions on the form of
$\xi_\mu$ inside the Schwarzschild horizon. In particular, the derivation of the
maximal redshift horizon in Sec.~\ref{subsec:Redshift} and the calculation
of the outbound photon in Sec.~\ref{subsec:Case2-outbound-photon-} are
independent of the $\xi_\mu$ configuration at  $r \leq 2M$.

Now add to the action \eqref{eq:actioncase2} the photon action from
\eqref{eq:LagMMgrav} using \eqref{eq:ansatznonbire}--\eqref{eq:ansatzxi}
and replace $\xi_\mu$ there by $\partial_\mu \phi$.
For $\phi=\overline{\phi}$ from \eqref{eq:overline-phi}, this then reproduces,
in the geometrical optics approximation, the \emph{Case--2} model
discussed in Sec~\ref{subsec:Case2} of the main text.\footnote{It
appears to be difficult to obtain the \emph{Case--3} model
as the solution of a pure scalar theory,
because its parameters $\xi^\mu$ cannot be written as the
gradient of a scalar field. If, however, we restrict ourselves to
spherically symmetric fields, we can use the scalar theory \eqref{eq:actioncase2}
also for the Case--3 model. The idea is to interpret the spherically
symmetric fields as belonging to a $(1+1)$--dimensional reduced theory
(possibly coming from a higher-dimensional gauge field theory~\cite{Witten1977})
and to use the Levi--Civita symbol $\epsilon^{mn}$ normalized by
$\epsilon^{01}=1$. Specifically, the dynamic Lorentz-violating
parameters in spherical coordinates are given by
$\xi^\mu\,\big|_\text{Case 3}=(\overline{\xi}^0,\, \overline{\xi}^1,\, 0,\, 0)$
with $\overline{\xi}^m = \epsilon^{mn}\,\partial_n \overline{\phi}$
for the gradient $(\partial_0,\, \partial_1)\equiv(\partial_t,\, \partial_r)$.
\label{ftn:case3-dynamics}}

The model \eqref{eq:actioncase2} just serves the purpose of a proof of
principle and shows that there can be dynamic Lorentz-violating theories
with the horizon structure discussed in this article.
It is certainly not a serious phenomenological model and may have
difficulties with causality and stability,
in addition to the obvious non-renormalizability. At the classical level,
though, where most of the considerations of our article apply,
it provides an example of a theory yielding one of the suggested background
configurations $\xi_\mu$.
\end{appendix}



\end{document}